\documentclass[prd,singlecolumn,aps,superscriptaddress,eqsecnum,floatfix]{revtex4-1}

\usepackage{empheq}
\usepackage{amsmath}
\usepackage{amssymb}
\usepackage{dsfont}
\usepackage{graphicx}
\usepackage{hyperref}
\usepackage{stackrel}
\usepackage{color}
\usepackage{comment}

\def \ee{\end{equation}}
\def \be{\begin{equation}}
\def \eea{\end{eqnarray}}
\def \bea{\begin{eqnarray}}

\usepackage[T1]{fontenc} 

\usepackage{amsmath,mleftright}
\usepackage{xparse}
\usepackage{mathtools}

\begin{document}


\title{ 
 A technique for natural gauge boson masses
}


\author{B. Koch}
\email{bkoch@puc.uc.cl}
\affiliation{Instituto de F\'isica, Pontificia Universidad Cat\'olica de Chile,\\Av. Vicku\~na Mackenna 4860, Santiago, Chile\\
Institute for Theoretical Physics, TU Wien, Wiedner Hauptstr. 8, A-1040 Vienna, Austria}
\author{C. Laporte}
\email{calaporte@uc.cl}
\affiliation{Instituto de F\'isica, Pontificia Universidad Cat\'olica de Chile,\\Av. Vicku\~na Mackenna 4860, Santiago, Chile}


\begin{abstract}
In this work, a novel mechanism for spontaneous symmetry breaking is presented.
This mechanism avoids quadratic divergencies and is thus capable of
  addressing the hierarchy problem in gauge theories. 
Using the scale-dependent effective action $\Gamma_{k}$ minimally coupled to a gravitational sector, variational parameter setting is applied.
This provides a mass and vacuum expectation value as a function of the 
constants arising in the low scale expansion of Newtons' and cosmological couplings. 
A comparison with experimental data, such as the Higgs mass, allows putting restrictions on these  
constants.
With this generic approach one can compare with explicit candidates for an effective field theory of gravity.
As an example, we use the asymptotic safety scenario, where we find restrictions on the matter content of the theory.
\end{abstract}


\maketitle

\tableofcontents
\section{Introduction}\label{sec:intro}
The Standard Model describes the elementary particles and their interactions in a successful way.  
However, there are good reasons for looking for physics beyond the Standard Model.
 One such motivation is the subject of this paper; the so-called hierarchy problem
 in theories with 
 Spontaneous Symmetry Breaking (SSB)~\cite{tHooft:1979rat,deGouvea:2014xba,Georgi:1974yf,Susskind:1978ms,Dine:2015xga,Giudice:2008bi}.

\subsection{SSB, quadratic divergencies, and the hierarchy problem in a nutshell}
\label{subsec:Intro1}

The measurements of the Higgs boson at the Large Hadron Collider~\cite{Aad:2012tfa, Chatrchyan:2012xdj} confirms that 
its existence and properties are consistent with the Standard Model (SM). 
Unlike all other particles of the  SM the Higgs is a fundamental scalar, which 
gives rise to  the question, whether
 the SSB mechanism, which is induced by the Higgs field, is natural~\cite{Vissani:1997ys,Foot:2007ay,Foot:2007iy,Grinbaum:2009sk,Wetterich:2011aa,Vieira:2012ex,Farina:2013mla,Heikinheimo:2013fta,Tavares:2013dga,Farzinnia:2013pga,Kobakhidze:2014afa,Haba:2014oxa,Craig:2015pha,Jegerlehner:2015cva,Pelaggi:2017wzr}. 
 The central issue is the strong sensitivity of masses of scalar particles under radiative corrections, 
 leading to the so-called hierarchy problem and the failure of the notion of naturalness.
It is, of course, possible that naturalness is not always a good guiding principle for the understanding
of nature \cite{Hossenfelder:2018ikr}, but if one would have the choice, a natural description
is certainly preferable.

The hierarchy problem affects only scalar particles  since Dirac and gauge fields are technically natural. 
Just as for scalar fields, the mass term associated with spin-1/2 or spin-1 fields is invariant under a global symmetry, but there is further an enhanced symmetry when the mass parameter goes to zero. 
Since, in this case, quantum corrections respect the enhanced symmetry, 
the  associated corrections will be proportional to the symmetry-breaking term.  
As a consequence, the loop corrections  to masses of  Dirac and gauge fields
will be suppressed by the smallness of the tree-level parameter, and no fine-tuning is needed when further corrections are incorporated.

Unfortunately, this does not hold for the mass term of scalar particles. In the SM, the term $m^2H^{\dagger}H$, with $H$ being the $SU(2)_{L}$ Higgs field doublet, is invariant under any gauge or global symmetry acting on it. The only symmetry acting on the scalar field when $m_{H}\rightarrow 0$ is the scale symmetry, however, at a mass scale $M$, much larger than the electroweak scale, the conformal symmetry is broken by $M$, as explained in~\cite{Giudice:2013yca}, and therefore the mass parameter is exposed to any contribution coming from the UV sector. 
This sensitivity to the quantum effects in the 
UV reflects the lack of arguments to justify the stability of the Higgs mass parameter against radiative corrections. 
Quite generally, in a theory with multiple mass-scales one finds that the Higgs mass $m$ 
accumulates quantum corrections from all (coupled) particles at all energy scales. 
Thus,
 $m$ is affected by heavy particles through
 the appearance of quadratic divergences (unlike the technically natural spin-$\frac{1}{2}$ or spin-1 fields). 
 Specifically, in an effective field theory approach of the SM, where the momenta of virtual particles are involved,  the radiative corrections are cut-off at the scale $\Pi$.
  The dependence of $m$ on
some experimental-scale $\mu$ can be obtained from~\cite{Veltman:1980mj},
 \begin{subequations}\label{hierarchy}
    \begin{gather}
    \label{hierarchy1}
    m^2_{H}\left(\mu\right)=m^2_{0}\left(\Pi\right) + \delta m^2,\\
    \label{hierarchy2}
    \delta m^2=\frac{\Pi^2}{16\pi}\left(\frac{3}{4}g^2_{1}+\frac{9}{4}g^2_{2}+3\lambda^2_{h}-12\lambda^2_{t}\right),
    \end{gather}
\end{subequations}
where $\lambda_{h},\lambda_{t},g_{1}$ and $g_{2}$ are the Higgs quartic, Yukawa, electroweak $SU(2)_{L}$, and hypercharge couplings, respectively. It is possible to 
provide a physical meaning for the cutoff~$\Pi$. For example, 
from a Wilsonian perspective, $\Pi$ would represent the space-time lattice spacing. 
Moreover, the quadratic divergences can be seen (at least approximately) as a placeholder for a physical threshold, 
for example heavy particles coupled to the Higgs. 
Following (\ref{hierarchy1}), an explanation of why the observed Higgs mass remains small requires a large fine-tuning such that the tree-level parameter exactly cancels the huge correction in (\ref{hierarchy2}). If one starts with a bare action, the quadratic divergences for scalar fields always arise 
when the quantum corrections are incorporated. 
Accordingly, a light Higgs scalar cannot survive in a natural way 
if the theory is expected  to hold  up to large energy scales, such as the Planck scale. This is referred to the fine-tuning, hierarchy, or naturalness problem and turns out to be independent of the scheme one uses to renormalize the theory. 

Historically speaking, there are three traditional ways of addressing the problem of quadratic divergences. The first one is embedding the SM into 
a new kind of symmetry, which acts in such a way that the Higgs mass is protected  by this symmetry, turning it into a technically natural parameter. 
One possible option to implement this  idea
is based on a new fermion-boson symmetry called supersymmetry~\cite{Wess:1973kz,Wess:1974tw,Fayet:1977yc,Witten:1981nf,Haber:1984rc}, where a cancellation between loops of different statistics takes place. 

The second option invokes the possibility of bringing down the cutoff of the SM through an electroweak symmetry breaking spawned by a dynamically generated vacuum condensate of a strongly coupled group, known as technicolor or Higgsless models~\cite{Weinberg:1975gm,Susskind:1978ms,Hill:2002ap,Csaki:2003dt,Csaki:2003zu}. Nevertheless, the discovery of a light Higgs mass~\cite{Aad:2012tfa, Chatrchyan:2012xdj} 
ruled most of these models out. However, various extensions~\cite{Goldberger:2008zz,Fan:2008jk,Martin:2008cd,Sannino:2009za,Chacko:2012vm} suggest alternative ways to preserve this idea.

The third option establishes a set of vacua of the SM, over which the Higgs mass varies according to some statistical distribution~\cite{Agrawal:1998xa,Tegmark:2003ug,Donoghue:2009me}. The anthropic principle provides a guide for explaining the observed light Higgs boson mass and the closeness to the QCD and weak scales without resorting to additional symmetries or a lower cutoff. 

An incomplete list of
more recent alternatives includes: \\
\textit{4) NNaturalness}~\cite{Arkani-Hamed:2016rle,Cohen:2018cnq}, which relies on multiples copies on the SM, each with a different vacuum expectation value. \\
\textit{5) Twin Higgs}~\cite{Chacko:2005pe,Barbieri:2015lqa,Low:2015nqa} models are based on the incorporation of discrete symmetries that allow different SM quantum numbers, and the representation of two Higgs doublets into a fundamental $SU(4)$. \\
\textit{6) Noncommutative perturbative dynamics}~\cite{Minwalla:1999px} assumes the separation of UV and IR physics, 
using a noncommutative theory. Here, non-trivial mixtures of UV and  IR phenomena can explain different hierarchies in nature. \\
\textit{7) Asymptotic Safety}~\cite{Shaposhnikov:2009pv,Eichhorn:2017als,Pawlowski:2018ixd,Kwapisz:2019wrl,Wetterich:2019rsn} 
works with a non-trivial UV fixed point for gravity. 
In this approach it was possible to predict the observed Higgs mass from
supposing that the Standard Model plus gravity are valid up to Planck scale energies and assuming that
there are no new fundamental degrees of freedom at intermediate scales. \\
\textit{8) Cosmological relaxation}~\cite{Graham:2015cka} is a model, where the cosmological evolution of the Universe 
is diving the Higgs boson mass to a much smaller value than the Planck scale.

\subsection{The idea of SSB without quadratic divergencies}

The goal of this paper is to point out a novel way of inducing spontaneous symmetry breaking.
It allows generating masses for gauge bosons without quadratic divergences.
This mechanism avoids the hierarchy problem mentioned in the previous subsection. 

The starting point will be a bare action
without ``dangerous'' interactions like the quartic Higgs coupling. 
Quantum corrections to this classical bare action yield a scale-dependence at the level of the gauge couplings contained in an effective action $\Gamma_{k}$. 
The arbitrary renormalization scale cannot be part of physical observables. It will be set 
following the Variational Parameter Setting (VPS)~\cite{Koch:2014joa} prescription, 
which can be understood as the
principle of minimal sensitivity~\cite{Stevenson:1981vj}, applied to quantum field theory background calculations. 
The VPS prescription allows minimizing
the scale-dependent effective action with respect to variations of its field- and parameter content.
Variations with respect to the field content give the well-known gap equations~\cite{Avan:1983bv},
which are complemented by one, variationally derived, scale setting equation.

It is possible to choose a bare action $S(\Phi)$ such that no 
quadratic divergences arise in the effective action $\Gamma_{k}(\phi)$,
where $\phi$ is the quantum expectation value of the field $\Phi$.
This means for constant values of the renormalization scale $k$, there will be no terms, which are usually necessary 
to generate SBB. 
However, in every quantum field theory calculation, the scale $k$ has to be set
in order to arrive at a testable prediction. This necessity is a consequence of the incomplete
nature of any perturbative or effective quantum field theory approach.
It depends on the observable one is interested in, whether one chooses $k$ as a function of kinematic variables,
renormalized parameters, or something else. Thus, assuming the scale to be an independent constant
was actually inconsistent.
For example, when one is interested in background configurations, one way to proceed
is the scale setting of the ``improving solutions''
procedure, leading to Uehling-type potentials~\cite{Uehling:1935uj,Dittrich:1985yb,Bonanno:1998ye,Wetterich:1992yh,Reuter:1993kw,Reuter:1992uk,Bonanno:2000ep,Litim:2007iu,Koch:2010nn,Contreras:2013hua,Koch:2013owa,Pagani:2019vfm}. 
However, the scale setting in ``improving solutions procedures'' leads to an anomalous violation of the underlying gauge symmetries. Fortunately, such a breaking of gauge symmetries is not necessary.
As shown in~\cite{Koch:2014joa}, the aforementioned 
VPS prescription allows to derive  an optimal scale setting $k\rightarrow k_{opt}$,
which preserves  the underlying gauge symmetries of the effective action $\Gamma_{k}$.
After the replacement $k\rightarrow k_{opt}$ the  effective action $\Gamma_{k}$ becomes an
optimal effective action $\Gamma_{opt}$.

The main massage of this paper is that for SSB to occur, it is sufficient that it occurs at the level
of this optimal effective action $\Gamma_{opt}$, as opposed to the non-optimal effective action $\Gamma_{k}$,
or the bare action $S$.
 The advantage of this is that quadratic divergences arising from quantum corrections of 
 the  bare action $S$ in the standard SSB 
 are absent when SSB only occurs at the level of $\Gamma_{opt}$, since the optimal effective action
 has all quantum corrections already incorporated~\cite{Reuter:2007rv, Rosten:2010vm}. No additional quantum 
 corrections have to be incorporated into $\Gamma_{opt}$ and  thus, no quadratic divergencies occur.

This idea is conceptually appealing. In section \ref{sec:two} it is implemented for scalar Quantum Electro Dynamics (QED).
It turns out that the resulting optimal effective action $\Gamma_{opt}$ only allows for
SSB if  the gauge fields form a condensate. 
Even though this might be an interesting possibility, it deviates from our original intention.
One realizes that for the program to work, one needs a scale-dependent vacuum contribution 
to the effective action.
 This is the reason why we proceed with an effective description of quantum gravity, where
 this vacuum contribution is given in terms of a cosmological constant.
 As an example, we study the Asymptotic Safety (AS) approach.
The AS~\cite{Hawking:1979ig} 
conjecture provides a consistent description of gravity as a non-perturbatively renormalizable quantum field theory~\cite{Niedermaier:2006ns,Niedermaier:2006wt,Percacci:2011fr} and a scenario for testing the results of this work. Moreover, the scale-dependence of the gravitational and cosmological constant has been extensively studied, in~\cite{Reuter:1996cp,Dou:1997fg,Reuter:2004nx,Christiansen:2012rx,Biemans:2016rvp,Wetterich:2017ixo,Canales:2018tbn} as well as properties and consequences of the scale-dependent Einstein-Hilbert action~\cite{Souma:1999at,Lauscher:2001ya,Reuter:2001ag,Litim:2003vp,Fischer:2006fz,Codello:2006in} and Gaussian massless matter fields minimally coupled to an external metric~\cite{Percacci:2002ie,Percacci:2003jz,Eichhorn:2011pc,Eichhorn:2012va,Dona:2013qba,Dona:2014pla,Dona:2015tnf,Eichhorn:2017egq}.

The paper is organized as follows: 
Below, a discussion of similarities and differences with the Coleman Weinberg (CW) mechanism~\cite{Coleman:1973jx}. 
In section~\ref{sec:two}, an optimized effective action is obtained for the electromagnetic sector in the low-energy behavior of the U(1) coupling and the anomalous dimension.  
In section~\ref{sec:three}, the gravitational sector is taken into account, 
conducing to a symmetry breaking potential.  
Section~\ref{sec:four} provides expressions for the mass and vacuum expectation value of the Higgs boson as well as a benchmark for gravitational parameters coming from an infrared expansion.
Section~\ref{sec:five} connects our results with those obtained in the context of the functional renormalization group. 
Finally, a summary, some comments, and ideas for future work are given in section~\ref{sec:six}.

\subsection{Similarities and differences with the Coleman Weinberg mechanism}

The model presented is strongly reminiscent of the CW mechanism~\cite{Coleman:1973jx} because neither of the two shows SSB at tree level. However, there are crucial 
differences, which are explored by outlining both methods.

\begin{itemize}
    \item In the CW mechanism, the radiative corrections drive the SSB in theories that do not exhibit such breakings at tree level. The theory considers a massless scalar QED where the radiative corrections generate a one-loop effective potential. The mass of the scalar meson is defined as the second derivative of the effective potential evaluated at the arbitrary mass scale $M$, introduced to stay away from the logarithmic singularity in momentum space. The identification of this scale is arbitrary, generically chosen, such that the masses of scalar and vector fields give the observed values. After SSB, the theory can be expressed in terms of one dimensionless $e$ and one dimensionful (the VEV $v$). Even though the mass parameter is absent, some phenomena indicate the naturalness problem persists. The original prediction for the Higgs mass ($\sim 10 GeV$) is far from the currently observed value, thus forcing the Higgs self-coupling to be so large that a nearby Landau pole might break the one-loop approximation~\cite{Meissner:2006zh}. Furthermore, if the cutoff scale is of TeV's order, the mass of the top quark is less than the observed value~\cite{Fatelo:1994qf, Chaichian:1995ef}.
    \item In the present model, we start with a scalar QED theory without quartic interaction, minimally coupled with the metric field. The main consequence of the proposed approach lies in the fact that the one loop quantum corrections do not generate SSB at the level of the effective action $\Gamma_{k}$. To see the previous statement, consider the one-loop effective potential generated in scalar QED~\cite{Coleman:1973jx}
    
    \begin{eqnarray}\label{scalarQEDep}
        V_{eff}[\phi]=m^2\phi^2+\frac{\lambda}{4!}\phi^4-\frac{1}{4(4\pi)^2}\left\{3\left(2e^2\phi^2\right)^2\left(\log{\frac{M^2}{2e^2\phi^2}}+\frac{3}{2}\right)+\right.\\
        \left.\left(m^2+\frac{\lambda \phi^2}{4}\right)^2\left(\log{\frac{M^2}{m^2+\frac{\lambda \phi}{4}}}+\frac{3}{2}\right)+\left(m^2+\frac{\lambda \phi^2}{12}\right)^2\left(\log{\frac{M^2}{m^2+\frac{\lambda \phi}{12}}}+\frac{3}{2}\right) \right\}\nonumber
    \end{eqnarray}
    
    where $M$ corresponds to an infrared cutoff introduced to avoiding the logarithmic infrared singularity. Since (\ref{eq:one}) does not have quartic interaction, we put $\lambda=0$ in (\ref{scalarQEDep}). The m-value can be inferred from the location of the minimum of $V_{eff}$ at the vev $\phi=\left\langle \phi \right\rangle$. After inserting the coupling $m$ in the effective potential, the coupling $\lambda$ is defined by,
    
    \begin{align}\label{lambdaeff}
        \lambda&=\frac{d^4V_{eff}(\left\langle \phi \right\rangle)}{d\phi^4}\nonumber\\
        &=\frac{27e^8\left(\log{\frac{3e^2}{32}}-1-2\log{\pi}\right)}{1024\pi^6}.
    \end{align}
    
    In the last step we have used $M=\sqrt{2}\mathrm{e}^{-\frac{1}{4}}e\left\langle \phi \right\rangle$, with $\mathrm{e}$ being the Euler constant. From (\ref{lambdaeff}) one can deduce that $\lambda$ is proportional to $e^2$, i.e., corrections of higher orders. However, by applying the variational parameter setting (\ref{eq:four}), one gets SSB at the level of the optimal effective action $\Gamma_{opt}$ due to the existence of a vacuum gravitational term, concluding that radiative corrections, together an appropriate way of fixing the renormalization scale are the dominant driving forces of SSB. Additionally, the unphysical dependence of the renormalization scale is removed from observables. As a consequence, the mass for the scalar and vector fields are functions of the parameters appearing in the original Lagrangian, which include the unknown low-energy expansion of gravitational couplings.
\end{itemize}

Thus, both approaches have in common the lack of SSB at the level of the classical action and the dependence of physical observables on an arbitrary parameter introduced to avoid infrared logarithmic singularities. However, they differ in the way of generating SSB. In the CW mechanism, radiative corrections are enough to ensure a mass term for the scalar field. In the present model, scalar QED without quartic interaction will be coupled to the gravitational sector, and the symmetry breaking potential will be driven by the scale setting condition (\ref{eq:four}). This will allow to address the Hierarchy problem in a way which is not possible in the CW formalism.

\section{Scalar QED without quartic interaction}\label{sec:two}

As first example one can consider a theory that contains charged
spin-zero particles that interact with photons. The bare action is given by,
\begin{equation}\label{eq:one}
\small{
\mathcal{S}(A_{\mu},\Phi)=\int d^4x \left(\frac{a_{b}}{2}(D_{\mu}\Phi)^*(D^{\mu}\Phi)+\frac{m^2_{b}}{2}\Phi^*\Phi-\frac{1}{4e^2_{b}}F_{\mu\nu}F^{\mu\nu}\right)},
\end{equation} 
where $D_{\mu}=\partial_{\mu}-iA_{\mu}$, $F_{\mu\nu}=\partial_{\mu}A_{\nu}-\partial_{\nu}A_{\mu}$, and $\Phi$ is a complex scalar field. 

\subsection{Optimized effective action}
\label{sec:twopointone}

Following Wilsons' idea~\cite{Wilson:1974mb}, one can define an average effective action $\Gamma_{k}$ as the functional obtained after integrating out the quantum fluctuations, which contain momenta $q^2 > k^2$. By changing k, this scale-dependent effective action can be seen as a smooth interpolation between the microscopic ultraviolet action $\Gamma_{k\rightarrow \infty}$ and the full quantum effective action in the infrared limit $\Gamma_{k\rightarrow 0}$. The effective action for (\ref{eq:one}) reads,
\begin{equation}
\label{eq:two}
\Gamma_{k}=\int d^4x\left(\frac{a_{k}}{2}(D_{\mu}\phi)^*(D^{\mu}\phi)+\frac{m_{k}}{2}\phi^*\phi-\frac{1}{4e^2_{k}}F_{\mu\nu}F^{\mu\nu}\right).
\end{equation}
This effective action has no hierarchy problem, but also no standard SSB.
The couplings ($a_{k},m_{k},e_{k}$) are now scale-dependent quantities. To avoid the logarithmic divergences appearing in the QED couplings due to deep infrared scale $k\rightarrow 0$, the Renormalizaton Group (RG) scale is 
split into its reference fixed part $k_{0}=m_{0}$ and its variable part $k^{'}$ as $k=m_{0}+k^{'}$. Identifying the reference scale as $m_{0}$ and defining the gravitational couplings in the infrared limit at this scale as shown in~\autoref{fig:referencescale}, the couplings can be expanded in the vicinity of $m_{0}$ using the dimensionless quantity $\frac{k'}{k_{0}}$ as the expansion parameter
\begin{subequations}\label{eq:three}
\begin{align}
\label{eq:threeone}
a_{k}&=a_{0}+\xi_{a,1}\frac{k^{'}}{m_0}+\xi_{a,2}\frac{{k^{'}}^2}{m^2_{0}}+\mathcal{O}\left(\frac{k^{'}}{m_{0}}\right)^3\\
\label{eq:threetwo}
m^2_{k}&=m^2_{0}+\xi_{m,1}\frac{k^{'}}{m_0}+\xi_{m,2}\frac{{k^{'}}^2}{m^2_{0}}+\mathcal{O}\left(\frac{k^{'}}{m_{0}}\right)^3\\
\label{eq:threethree}
\frac{1}{e^2_{k}}&=\frac{1}{e^2_{0}}+\xi_{e,1}\frac{k^{'}}{m_{0}}+\xi_{e,2}\frac{{k^{'}}^2}{m^2_{0}}+\mathcal{O}\left(\frac{k^{'}}{m_{0}}\right)^3.
\end{align}
\end{subequations}
The set of coefficients ($\xi_{ij}$) with $i=(a,m,e)$ and $j=(1,2)$ is obtained from the beta functions of (\ref{eq:one}). Those beta functions further depend on the renormalization scheme.

It is desirable that physical observables are independent of the particular renormalization scheme used to renormalize the theory and the corresponding unphysical parameters involved in this process. However, if  
the prediction, calculated by a series of approximations,
depends on unphysical parameters, then the parameters should be chosen such that 
variations will minimize the sensitivity of the observable on those parameters. 
Following this criterion, one looks for a scale setting of the renormalization scale as a function of physical variables $k=k(\phi,\xi_{i},F_{\mu\nu},...)$. This identification results from applying the variational principle to 
$k$ by promoting the scale $k$ to a field \cite{Koch:2014joa} at the level of the effective action.
As shown in~\cite{Koch:2010nn, Domazet:2012tw, Koch:2014joa}, this minimization can be written as
\begin{eqnarray}\label{eq:four}
        &&\frac{\delta \Gamma\left(A_{\mu},\phi(x),k(x),a_{k},m_{k}\right)}{\delta k}=0\\
        \Rightarrow && \frac{\text{d}\mathcal{L}\left(A_{\mu},\phi(x),k(x),a_{k},m_{k}\right)}{\text{d}k} \Bigr|_{k=k_{opt}}=0.\nonumber
\end{eqnarray}
In contrast to most other scale settings, the above procedure allows maintaining the original gauge symmetries
(even diffeomorphism invariance \cite{Koch:2014joa}).
The philosophy underlying this procedure has been developed in~\cite{Koch:2010nn, Domazet:2012tw, Koch:2014joa}
and it has been successfully applied in different contexts~\cite{Morris:2016spn, Koch:2016uso, Rincon:2017ypd, Rincon:2017goj, Rincon:2018lyd, Platania:2019qvo}.
For consistency, with the expansion (\ref{eq:three}), also the effective action has been expanded up to $k^2$. 
The condition (\ref{eq:four}) allows resolving the renormalization-point ambiguity by selecting a single scale and fixing 
it as a function of dynamical field variables,
\begin{eqnarray}\label{eq:five}
k_{opt}\rightarrow\frac{-\xi_{e,1}F_{\mu\nu}F^{\mu\nu}+2\xi_{a,1}\left|D_{\mu}\phi\right|^2-2\xi_{m,1}\phi^2}{2\left(\xi_{e,2}F_{\mu\nu}F^{\mu\nu}-2\xi_{a,2}\left|D_{\mu}\phi\right|^2+2\xi_{m,2}\phi^2\right)}.
\end{eqnarray}
Since there are no new undetermined integration constants in (\ref{eq:five}), one can
insert the solution $k=k_{opt}$ back into (\ref{eq:two}).
This gives an optimal effective action independent of the arbitrary scale $k$,

\begin{eqnarray}\label{eq:six}
\Gamma_{opt}=\int d^4x \left\{ \frac{2\xi_{e,1}\xi_{e,2}\xi_{m,1}-4m^2_{0}\xi^2_{e,2}-\xi^2_{e,1}\xi_{m,2}}{8\xi^2_{e,2}}\phi^2+ \right. \\
                        \left. \frac{\left(\xi_{e,2}\xi_{m,1}-\xi_{e,1}\xi_{m,2}\right)^2}{4F^{\mu\nu}F_{\mu\nu}\xi^3_{e,2}}\phi^4 + \mathcal{L}_{kin}+\mathcal{L}_{const}+\mathcal{O}(\phi^5) \right\}
                        \nonumber,
\end{eqnarray}
where the potential has been expanded to order $\phi^4$ in a weak field approximation. 
Kinetic factors of $\phi$ and $A_\mu$  are contained in $\mathcal{L}_{kin}$ and quantities without any field factor $\phi$ are assembled into $\mathcal{L}_{const}$. One notes that (\ref{eq:six}) has a quadratic and a quartic term in $\phi$, which is
necessary for the standard SSB mechanism to take place.
It is important to remember that the effective actions (\ref{eq:two} and \ref{eq:six}) 
contain already quantum corrections. Thus, the quantization of the quartic interaction $\sim \phi^4$ 
in (\ref{eq:six}) does
not introduce any hierarchy problem, because it is already quantized.
 However, one also notes that the quartic coupling is only well behaved for a finite electromagnetic background field, $<F_{\mu\nu}F^{\mu\nu}>\neq 0$. Even though this is an interesting feature, it is not the type of SSB we are interested in. 

\subsection{Values of parameter expansion from QED sector}\label{twopointtwo}

\begin{figure}[!h]
\centering
\includegraphics[width=0.8\linewidth]{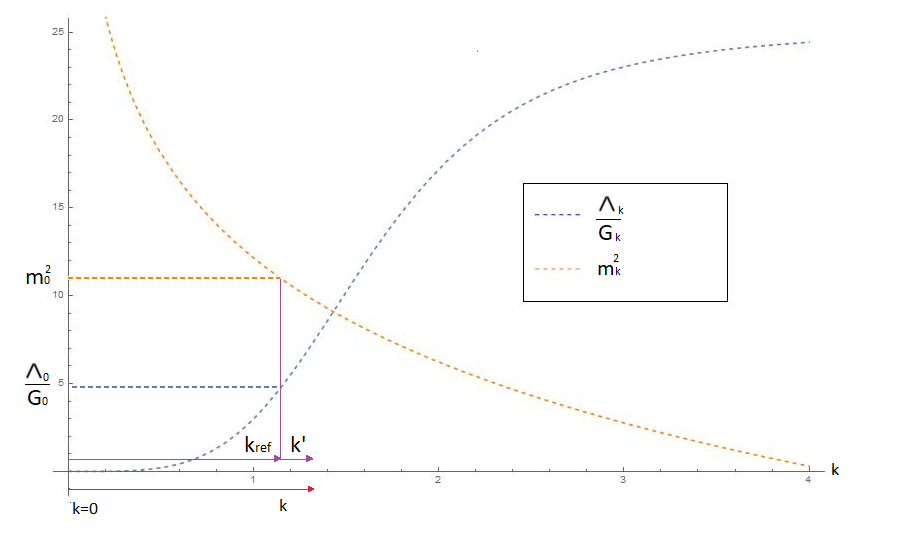}
\caption{Graphic representation of reference scale used in this work.}
\label{fig:referencescale}
\end{figure}

It is instructive to calculate the explicit values of the parameters $\xi_{e,j},\xi_{m,j}$. 
Those $\xi$s have some scheme dependence and can be obtained by applying perturbative methods. When the integral for obtaining an explicit expression of the coupling as a function of the scale is carried out, the lower limit (unlike the case of gravity) cannot be zero, which explains the expansion around $k_{0}$ instead of 0 in (\ref{eq:three}). Expressing $k\frac{\partial \alpha}{\partial k}$, with $\alpha \equiv \frac{e^2}{4\pi}$, in terms of the known scalar QED $\beta$-function up to one loop in the minimal subtraction scheme~\cite{Peskin:1995ev} gives the RG equation.

\begin{equation}
    \label{eq:sevenone}
    k\frac{\text{d}\alpha}{\text{d}k}=\frac{\alpha^2}{6\pi}.
\end{equation}
Integrating this equation  between the initial and an intermediate scale, 
the running coupling  takes the form,  

\begin{equation}
\label{eq:seven}
\alpha(k^2)=\frac{\alpha(k_{0})}{1- \frac{1}{3\pi}\alpha(k_{0})\text{ln}\left(\frac{k}{k_{0}}\right)}.
\end{equation}
The expansion for the running coupling $e_{k}$ around $k=k_{0}$ by imposing $e_{k} \rightarrow e_{k_0} \equiv e_{0}$ for the first term of the series (see Figure~\ref{fig:referencescale}) and rearranging (\ref{eq:seven}) gives,
\begin{equation}
\label{eq:eight}
\frac{1}{e^2_{k}} = \frac{1}{e^2_{0}}-\frac{1}{24\pi^2}\left(\frac{k^{'}}{m_{0}}\right)+\frac{1}{48
\pi^2}\left(\frac{{k^{'}}^2}{m^2_{0}}\right)+\mathcal{O}\left(\frac{k^{'}}{m_{0}}\right)^3.
\end{equation}
The one-loop contribution to the anomalous mass dimension in Lorentz gauge~\cite{Peskin:1995ev}, $\gamma_{m}=-\frac{3e^2}{16\pi^2}$, follows the same treatment. 
Integrating $\gamma_{m}$ with the initial condition in $k_{0}$, an expansion (like (\ref{eq:eight})) for the running coupling $m_{k}$ is obtained,
\begin{equation}\label{eq:nine}
m^2_{k}=m^2_{0}-\frac{3e^2_{0}m^2_{0}}{8\pi^2}\left(\frac{k^{'}}{m_{0}}\right)+m^2_{0}e^2_{0}\left(\frac{24\pi^2+e^2_{0}(9\pi-1)}{128\pi^3}\right)\left(\frac{{k^{'}}}{m_{0}}\right)^2+\mathcal{O}\left(\frac{k^{'}}{m_{0}}\right)^3.
\end{equation}
Thus, the set of parameters ($\xi_{i,j}$) from the scalar sector can be identified from a comparison between (\ref{eq:three}), (\ref{eq:eight}) and (\ref{eq:nine}). One finds

\begin{subequations}\label{eq:ten}
    \begin{align}
        \label{eq:ten-one}
        \xi_{e,1}&=-\frac{1}{24\pi^2}\\
        \label{eq:ten-two}
        \xi_{e,2}&=\frac{1}{48\pi^2}\\
        \label{eq:ten-three}
        \xi_{m,1}&=-\frac{3e^2_{0}m^2_{0}}{8\pi^2}\\
        \label{eq:ten-four}
        \xi_{m,2}&=m^2_{0}e^2_{0}\left(\frac{24\pi^2+e^2_{0}(9\pi-1)}{128\pi^3}\right).
    \end{align}
\end{subequations}
The set of equations (\ref{eq:ten}) is valid only for scalar QED up to one loop in perturbation theory in Lorentz gauge, and
in the minimal subtraction scheme. Possible 
contributions on these U(1) coefficients coming from the gravitational sector will be discussed in section~\ref{sec:four}.

\section{Gravitational sector minimally coupled to a charged scalar}\label{sec:three}

Since the field theoretical content in (\ref{eq:one}) was not providing all 
ingredients needed for SSB, one needs to take into account some generalization.
As explained in the introduction, a cosmological term (vacuum energy density) 
is important for the implementation of our ideas.
Thus, let's consider a gravitational sector coupled to matter.
In the leading order truncation (meaning that higher-order dipheomorphism invariant operators like $R^2$, $R_{\mu\nu}R^{\mu\nu}$,... are neglected), the simplest effective action of gravity coupled to a charged scalar reads,
\begin{equation}\label{eq:eleven}
\Gamma_{k}=\int d^4x \sqrt{-g}\left\{\kappa(2\Lambda_{k}-R)+\frac{m^2_{k}}{2}\phi^{*}\phi \right.
        \left. +\frac{a_{k}}{2}(D_{\mu}\phi)^{*}(D^{\mu}\phi)-\frac{1}{4e^2_{k}}F^{\mu\nu}F_{\mu\nu}+c_f {\mathcal{L}}_{kin,\psi}\right\}.
\end{equation}
One notes that all couplings $(\kappa_k, \, \Lambda_k, \, m_k, \, a_k,\, e_k)$, except the one of the sterile fermions $c_f$, are scale dependent quantities. In this action, $\Lambda_{k}$ stands for the scale-dependent cosmological constant, $R$ is the Ricci scalar,
and $\kappa=\left(16\pi G_{k}\right)^{-1}$, where
 $G_{k}$ is the running counterpart of the gravitational coupling. 
 In addition to (\ref{eq:three}), an expansion around the infrared zone $k \rightarrow k_{0}$ in the gravitational coupling  
 is needed to get an optimal scale.
In order to maintain consistency with the expansion (\ref{eq:three}), the effective action (\ref{eq:eleven}) and the gravitational couplings also are expanded to the same order. We examine the solutions of the gravitational couplings with an Einstein-Hilbert truncation in the deep infrared. In this limit, one finds, independent of the implementation of the Wilsonian renormalization procedure, a RG running of Newtons'  and the cosmological constant of the form,
\begin{eqnarray}\label{eq:twelve}
G(k)&=&G_{0}\left(1+C_{1}G_{0}k^2+C_{2}G^2_{0}k^4\right)+\mathcal{O}(k^6)\\
\Lambda(k)&=&\Lambda_{0}+C_{3}\Lambda_{0}G_{0}k^2+C_{4}\zeta\left({G_{0}\Lambda_{0}}\right)k^4+\mathcal{O}(k^6),
\end{eqnarray}
with $C_{1,2,3,4}$ being real numbers. 
Depending on the sign of $C_1$, (\ref{eq:twelve}) shows screening or anti-screening property of gravity. 
When an expansion is made around $k_{0}$ instead of zero, 
one can redefine the constants $C_i$
giving
\begin{equation}\label{eq:thirteen}
\tilde{\Lambda}_{k}=\tilde{\Lambda}_{0}+\xi_{\tilde{\Lambda},1}m_{0}\left(\frac{k^{'}}{m_{0}}\right)+\xi_{\tilde{\Lambda},2}m^2_{0}\left(\frac{k^{'}}{m_{0}}\right)^2+\mathcal{O}(k^{'})^3,
\end{equation}
where a change of the cosmological variable  $\tilde{\Lambda}_{k}=\frac{\Lambda_{k}}{G_{k}}$ has been applied. 
Please note that if one is interested in the effective Higgs potential, (\ref{eq:thirteen}) is enough to get an overview of the gravitational contribution to this model. 
In particular,  $G_{k}$ from (\ref{eq:twelve}) is not needed, because this part of the action is proportional to $R$, which does not take part in the SSB process.
Based on a study on field-parametrization dependence of the renormalization group flow in the vicinity of non-Gaussian fixed points in quantum gravity, a beta function derived from EH action can be used to fix the parameters $\xi_{i}$ in (\ref{eq:thirteen}). 
One can further look at how the massless-matter fields affect Asymptotically Safe quantum gravity. 
In this case, the parameters $\xi_{i}$ would have a dependence on the number and the nature of matter fields. For now, the gravitational parameter set $\xi_{\tilde{\Lambda,j}}$ is kept arbitrary. In the next section, a physical benchmark for this set $\xi_{\tilde{\Lambda},j}$ will be worked out.

As in (\ref{eq:five}), the scale setting is performed by demanding  (\ref{eq:eleven}) to 
be insensitive under infinitesimal changes of $k$, giving

\begin{equation}\label{eq:fourteen}
k_{opt}=\frac{\mathcal{H}_{R,1}+\mathcal{H}_{F,1}+\mathcal{C}_{1}-2G^3_{0}\left(\xi_{m,1}\phi^2-|D\phi|^2\xi_{a,1}\right)}{\mathcal{H}_{R,2}+\mathcal{H}_{F,2}+\mathcal{C}_{2}+4G^3_{0}\left(\xi_{m,2}\phi^2-\xi_{a,2}|D\phi|^2\right)}.
\end{equation}
Herein, the functions $\mathcal{H}_{R}$ and $\mathcal{H}_{F}$ have as arguments the Ricci tensor $R$ and electromagnetic tensor $F_{\mu\nu}F^{\mu\nu}$, respectively, in addition to the infrared value of the running $G_{k}$. 
The constants $\mathcal{C}_{1,2}$ only contain infrared couplings and electromagnetic constants $\xi_{e,m,...}$. 
When the optimal scale is inserted back into the effective action (\ref{eq:eleven}) (that includes gravitational effects), one gets an optimal effective action independent of $k$,

\begin{eqnarray}\label{eq:fifteen}
\Gamma_{opt}=\int d^4x\sqrt{-g}\left\{+\frac{\mu^2}{2}|\phi|^2-\frac{\lambda}{4}|\phi|^4-\frac{1}{4\tilde{e}}F_{\mu\nu}F^{\mu\nu} \right.\\
        \left.+\mathcal{L}_{kin}+\mathcal{L}_{coup}+\mathcal{L}_{const}+\mathcal{O}(\phi^6) \right\}, \nonumber 
\end{eqnarray}
where  $\mathcal{L}_{kin}$ contains kinetic terms for the real scalar and gauge fields. Couplings with higher-order factors and Ricci scalar quantities are collected into $\mathcal{L}_{coup}$, and the Lagrangian part independent of Ricci scalar, electromagnetic strength and scalar field is named $\mathcal{L}_{const}$. The effective potential has again been expanded up to order $\phi^4$. The optimal effective action~(\ref{eq:fifteen}) is written following the usual notation with $\mu$ and $\lambda$ appearing in the Abelian Higgs mechanism,

\begin{subequations}\label{eq:sixteen}
    \begin{align}
        \label{eq:sixteen:one}
        \frac{\mu^2}{2}&=\frac{m^2_{0}}{2}+\frac{\xi_{\tilde{\Lambda},1}\left(\xi_{m,2}\xi_{\tilde{\Lambda},1}-2\xi_{m,1}\xi_{\tilde{\Lambda},2}\right)}{8\xi^2_{\tilde{\Lambda},2}}
        \\
        \label{eq:sixteen:two}
        \frac{\lambda}{4}&=\frac{\left(\xi_{m,2}\xi_{\tilde{\Lambda},1}-\xi_{m,1}\xi_{\tilde{\Lambda},2}\right)^2}{32\xi^3_{\tilde{\Lambda},2}}.
    \end{align}
\end{subequations}
If those parameters have the correct values, then the field $\phi$ acquires a vacuum expectation value (VEV), and the U(1) global symmetry will be spontaneously broken. Thus, (\ref{eq:fifteen}) shows that, even if one starts with a model like (\ref{eq:eleven}), which has not SSB, it is possible to get this feature for the optimal effective action $\Gamma_{opt}$.
This is particularly true if $\mu^2,\lambda > 0$. 



\section{Gauge boson masses}\label{sec:four}

In this section, restrictions on the RG parameters $\xi_i$ will be studied.

\subsection{Mass and vacuum expectation value of scalar and gauge fields}

The parameters $\xi_{m,1}$ and $\xi_{m,2}$ appearing in (\ref{eq:sixteen}) 
can be obtained studying the changes of the anomalous mass dimension. 
Since the action (\ref{eq:eleven}) considers the Einstein-Hilbert contribution, 
the gravitational sector may, in principle, have an impact on the behavior of electromagnetic couplings. 
Gravitational corrections to the beta function in quantum field theories have been analyzed in~\cite{Robinson:2005fj,Ebert:2007gf,Tang:2008ah,Toms:2008dq,Ebert:2008ux,Rodigast:2009zj,Zanusso:2009bs,Anber:2010uj,Toms:2011zza,Narain:2013eea}. 
Non-Abelian gauge fields coupled to gravity in (3+1) dimensions give rise to an 
additional term in the one-loop beta function proportional to the inverse square of Planck scale, improving the asymptotic freedom of N$=4$ Super Yang-Mills theory~\cite{Grisaru:1980nk,Mandelstam:1982cb}. In \cite{Toms:2010vy} it is pointed out that quadratic divergences coming from the gravitational sector are responsible for asymptotic freedom of QED beta function in a gauge-independent context with energy scale near the Planck scale. 
For the case where a complex scalar field is minimally coupling to perturbative quantized Einstein gravity with an explicit gauge dependence in the photon and graviton propagator, the total vacuum polarization tensor depends on the gauge parameters, surface terms, a dimensionless constant, and the ultraviolet momentum cutoff, as explained in~\cite{Felipe:2011rs,Felipe:2012vq}. 
In the last case, there are several reasons for neglecting the gravitational contribution to the usual beta function of the U(1) gauge coupling:

\begin{itemize}
    \item Choosing the gauge parameter, $\xi$ appearing in the graviton propagator in \cite{Felipe:2012vq}, equal to $\frac{5}{13}$, a cancellation of a gravitational contribution to be takes place.
    \item Using dimensional regularization instead of a momentum cutoff, such that the arbitrary parameter contained in the gravitational term is set to 0 \cite{Felipe:2011rs}.
    \item  Some studies~\cite{Pietrykowski:2006xy,Toms:2007sk,Charneski:2013zja} have shown the beta function of scalar electrodynamics possesses no contribution coming from gravitational interactions.
    \item In the  infrared $k\approx k_0$ all gravitational contributions to the beta functions of matter will be strongly suppressed
    by the Planck-scale. This is the reason why the standard model without gravity is a successful quantum field theory in the first place.
    \end{itemize}
Given the arguments expressed above, the gravitational contribution to the electromagnetic beta functions will be neglected.
One condition on the effective potential in (\ref{eq:fifteen}) for producing positive Higgs parameters and then SSB 
was that $\lambda >0$. This determines the sign of $\xi_{\Tilde{\Lambda},2}$.
 This can be seen by replacing (\ref{eq:ten-three}) and (\ref{eq:ten-four}) in (\ref{eq:sixteen:two}) and demanding $\lambda>0$. For a negative value of $\xi_{\Tilde{\Lambda},2}$ one has to solve the inequality
\begin{equation}\label{eq:seventeen}
\left[e^4_{0}\left(9\pi-1\right)\xi_{\tilde{\Lambda},1}+24e^2_{0}\pi^2\left(\xi_{\Tilde{\Lambda},1}+2m^2_{0}\xi_{\Tilde{\Lambda},2}\right)\right]^2<0,
\end{equation}
which has no solution for $e_{0},\xi_{\Tilde{\Lambda},1}\in {\rm I\!R}$. Thus,
 the requirement for the field $\phi$ to acquire a VEV is $\xi_{\Tilde{\Lambda},2}>0$. 
 From (\ref{eq:fifteen}) and (\ref{eq:sixteen}), the VEV of the scalar field is,
\begin{equation}\label{ec:eighteen}
\phi^2_{VEV}=\frac{2\left(4m^2_{0}\xi^2_{\tilde{\Lambda},2}+\xi_{\tilde{\Lambda},1}\left(\xi_{m,2}\xi_{\tilde{\Lambda},1}-2\xi_{m,1}\xi_{\tilde{\Lambda},2}\right)\right)\xi_{\tilde{\Lambda},2}}{\left(\xi_{m,2}\xi_{\tilde{\Lambda},1}-\xi_{m,1}\xi_{\tilde{\Lambda},2}\right)^2}.
\end{equation}
Suppose that the scalar potential in (\ref{eq:fifteen})  is near one of the minima (say the positive one), then it is convenient to define a fluctuation of the scalar field $\phi(x)=\phi_{VEV}+\eta(x)$. The squared mass of the complex scalar field $\eta(x)$ is then

\begin{equation}\label{eq:nineteen}
m^{2}_{\eta}=2\left(m^2_{0}+\frac{\xi_{\tilde{\Lambda},1}\left(\xi_{m,2}\xi_{\tilde{\Lambda},1}-2\xi_{m,1}\xi_{\tilde{\Lambda},2}\right)}{4\xi^2_{\tilde{\Lambda},2}}\right).
\end{equation}
One notes that unlike in the usual approach (\ref{hierarchy1}, \ref{hierarchy2}),
the mass (\ref{eq:nineteen}) does not depend on the gauge couplings and their expansion parameters.
Further, the
optimal  U(1) coupling $\tilde e$ in equation (\ref{eq:fifteen}) is shifted from 
its infrared value  $e_0$
\begin{equation}\label{eq:twenty}
\frac{1}{\Tilde{e}^2}=\frac{1}{e_{0}^2}+\frac{\left(4\xi_{\tilde{\Lambda},1}+\xi_{m,1}\phi^2_{VEV}\right)\left(4\xi_{e,2}\xi_{\tilde{\Lambda},1}-8\xi_{e,1}\xi_{\tilde{\Lambda},2}+h\phi^2\right)}{4\left(4\xi_{\tilde{\Lambda},2}+\xi_{m,2}\phi^2\right)^2},
\end{equation}
where $h=\xi_{e,2}\xi_{m,1}-2\xi_{e,1}\xi_{m,2}$. The mass term for the gauge bosons is obtained through the product between the inverse of (\ref{eq:twenty}) and the VEV of the scalar field (\ref{ec:eighteen}),

\begin{equation}\label{eq:twenty-one}
m^2_{A}=\frac{64\xi^3_{\Tilde{\Lambda},2}\left(\frac{\mathcal{F}_{1}}{4\xi^2_{\Tilde{\Lambda}}}-m^2_{0}\right)}{\left(\xi_{m,2}\xi_{\Tilde{\Lambda},1}-\xi_{m,1}\xi_{\Tilde{\Lambda,2}}\right)^2}\left(\frac{4}{e^2_{0}}+\frac{\mathcal{F}_{3}\left(\xi_{e,2}\mathcal{F}_{3}+6\xi_{e,1}\xi_{m,2}\xi_{\Tilde{\Lambda},2}\mathcal{F}_{1}-4\xi_{e,1}\mathcal{F}_{2}\right)}{\left(3\xi^2_{m,2}\xi^2_{\Tilde{\Lambda},1}\xi_{\Tilde{\Lambda},2}-6\xi_{m,1}\xi_{m,2}\xi_{\Tilde{\Lambda},1}\xi^2_{\Tilde{\Lambda},2}+2\mathcal{F}_{2}\right)^2}\right)^{-1},
\end{equation}

with, 

\begin{subequations}\label{eq:twenty.two}
    \begin{gather}
        \label{eq:twenty-two:one}
        \mathcal{F}_{1}=\xi_{\Tilde{\Lambda},1}\left(2\xi_{m,1}\xi_{\Tilde{\Lambda},2}-\xi_{m,2}\xi_{\Tilde{\Lambda},1}\right)
        \\
        \label{eq:twenty-two:two}
        \mathcal{F}_{2}=\xi^3_{\Tilde{\Lambda},2}\left(\xi^2_{m,1}+2m^2_{0}\xi_{m,2}\right)
        \\
        \label{eq:twenty-two:three}
        \mathcal{F}_{3}=2\xi^2_{m,2}\xi^3_{\Tilde{\Lambda},1}-3\xi_{m,1}\xi_{m,2}\xi^2_{\Tilde{\Lambda},1}\xi_{\Tilde{\Lambda},2}+4m^2_{0}\xi_{m,1}\xi^3_{\Tilde{\Lambda},3}.
    \end{gather}
 \end{subequations}

After insertion of the set of U(1) parameters (\ref{eq:ten}) into (\ref{ec:eighteen}) and (\ref{eq:nineteen}), the
mass and VEV of Higgs field are determined as a function of gravitational parameters appearing in the infrared expansion of $\frac{\Lambda_{k}}{G_{k}}$ and $m_{0}$, 

\begin{subequations}\label{eq:twentythree}
    \begin{gather}
        \label{eq:twentythreeone}
        v^2=256\pi^3\xi_{\tilde{\Lambda},2}\left(\frac{e^4_{0}\left(9\pi-1\right)\xi^2_{\tilde{\Lambda},1}+512m^2_{0}\pi^3\xi^2_{\tilde{\Lambda},2}+24e^2_{0}\pi^2\xi_{\tilde{\Lambda},1}\zeta_{1}}{e^4_{0}\left(e^2_{0}\left(9\pi-1\right)\xi_{\tilde{\Lambda},1}+24\pi^2\zeta_{2}\right)^2}\right),
        \\
        \label{eq:twentythreetwo}
        m^2_{\eta}=\frac{1}{256\pi^3\xi^2_{\tilde{\Lambda},2}}\left(e^4_{0}\left(9\pi-1\right)\xi^2_{\tilde{\Lambda},1}+512\pi^3m^2_{0}\xi^2_{\tilde{\Lambda},2}+24e^2_{0}\pi^2\xi_{\tilde{\Lambda},1}\zeta_{1}\right),
    \end{gather}
\end{subequations}
with $\zeta_{1}=\left(\xi_{\tilde{\Lambda},1}+4m_{0}\xi_{\tilde{\Lambda},2}\right)$ and $\zeta_{2}=\left(\xi_{\tilde{\Lambda},1}+2m_{0}\xi_{\tilde{\Lambda},2}\right)$. 
At this point, it is important to mark a few comments about the result obtained in (\ref{eq:twentythree}). 
The initial action (\ref{eq:fifteen}) has no elements to produce mass for the U(1) gauge boson since the quartic interaction is missing. After applying the VPS procedure, to get an optimal effective action, a symmetry breaking effective potential appears when the gravitational sector is taken into account.
 However, the RG expansion parameters, which drive this SSB must  meet certain  requirements. 
 The hermicity of the optimal Lagrangian implies that the parameters in (\ref{eq:sixteen}) must be real, and thus this Lagrangian respects charge, parity, and time-reversal symmetries. 
 Mimicking the usual Higgs mechanism for SSB, the sign of the mass term is chosen negative. 
 Moreover, the effective coupling $\lambda$ must be positive as a requirement for the scalar potential to be bounded from below. As shown above, this implies that $\xi_{\tilde{\Lambda},2}>0$.

\subsection{Benchmark of gravitational parameters}

Due to the dependence on infrared coefficients $\xi_{i}$ provided in the expansion of the couplings involved in the theory, one can expect to get restrictions from physically observed gauge boson masses. This exercise is done despite 
the fact that the model presented in (\ref{eq:fifteen}) is more a conceptual case study rather than a competitive phenomenological model since it has neither electro-weak nor Yukawa couplings implemented. 
Observed experimental values of the Higgs mass and the VEV will be employed to get a better idea about the distribution of the allowed parameters $\xi_{\Tilde{\Lambda},1}$, $\xi_{\Tilde{\Lambda},2}$ in agreement with the observations. The quantities $m_{\eta}$ and $v$ in (\ref{eq:twentythree}) have four free parameters, $e_{0}$, $m_{0}$, $\xi_{\Tilde{\Lambda},1}$ and $\xi_{\Tilde{\Lambda},2}$. We impose that,

\begin{itemize}
    \item $e_{0}=\sqrt{4\pi \alpha} \approx 0.3028$~\cite{Mohr:2015ccw}. 
    The U(1) coupling $e_{0}$ in (\ref{eq:eleven}) takes the value of the vertex function in spinor electrodynamics when all three particles (one incoming fermion, incoming photon and outgoing fermion)
    are on shell, i.e. the elementary charge $e$. In the deep infrared, the choice $k^{'} \rightarrow 0$ is justified due to the long-range character of~QED. 
    One confirms numerically that the difference between $\tilde e$ and $e_0$ is negligible.
    
     \item $v=\left(\sqrt{2}G_{F}\right)^{-\frac{1}{2}}=246.2197$~GeV \cite{Webber:2010zf}, because the experimental uncertainty on the Higgs mass $m_{H}$ is much larger than the uncertainty on the VEV of the Higgs field measured in the muon decay $v_{H}$, only the best fit value $v=246.2197$ will be considered to fix 
     $m_{0}$ as a function of the gravitational parameters $\xi_{\Tilde{\Lambda},1}$ and $\xi_{\Tilde{\Lambda},2}$.
    \item $ \xi_{\tilde{\Lambda},2}>0 $. When the two preceding points are applied to (\ref{eq:twentythreetwo}), the 
     bound on $\xi_{\Tilde{\Lambda},2}$ arises from imposing real values for  $m_{\eta}$.
\end{itemize}

\begin{figure}
    \centering
    \includegraphics[scale=1.25]{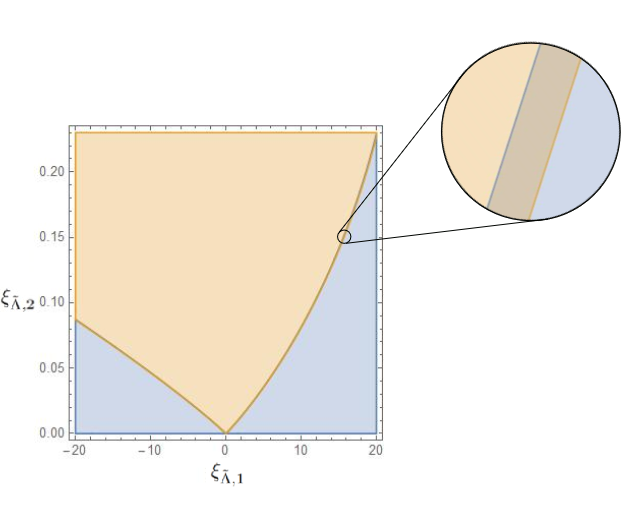}
    \caption{Allowed parameter range in the gravitational parameter $\xi_{\Tilde{\Lambda},1}$ and $\xi_{\Tilde{\Lambda},2}$. The orange region represents gravitational parameters which give a mass for the U(1) gauge boson lower than $m_{\eta}=125.33$ GeV while the blue region represents parameters greater than $m_{\eta}=124.85$ (lower and upper experimental limit of the mass of the Higgs boson).}
    \label{fig:two}
\end{figure}

The boundaries for the gravitational parameters $\xi_{\Tilde{\Lambda},1}$ and $\xi_{\Tilde{\Lambda},2}$ can be obtained by associating the limits of the Higgs boson mass with the limits of $m_{\eta}$ in (\ref{eq:twentythreetwo}).
The result is shown in \autoref{fig:two}. 
Parameters enclosed in the shadow region meet the experimental requirements previously discussed.

\section{Comparison with the functional renormalization group}\label{sec:five}

Up to now, our calculations never made use of a specific shape of gravitational beta functions. 
Let's make up leeway.
The evolution of the scale-dependent dimensionless couplings is dictated by the functional renormalization group equation ($a.k.a.$ Wetterich equation),

\begin{equation}\label{eq:twentyfour}
    \frac{\text{d}\Gamma_{k}}{\text{d}t}=\frac{1}{2}\text{STr}\left(\frac{\partial_{t}\mathcal{R}_{k}}{\Gamma^{(2)}_{k}[\phi]+\mathcal{R}_{k}}\right).
\end{equation}

The Wetterich equation is formulated such that it depends on renormalization group time $t=\ln{\frac{k}{k_{0}}}$, the modified inverse propagator $\Gamma^{(2)}_{k}[\phi]+\mathcal{R}_{k}$ involving a second functional derivative of the scale-dependent effective action with respect to the fields, and the momentum cutoff $\mathcal{R}_{k}$.
The cutoff is chosen such that it suppresses the contributions of field modes smaller than the cutoff scale $k^2$~\cite{Wetterich:1992yh,Morris:1993qb,Delamotte:2007pf,Gies:2006wv}. The notation $\text{STr}$ 
stands for ``Super-Trace'', which
is performed over momenta as well as particle species and spacetime or internal indices, 
including a factor $(-1)$ for fermionic fields. Despite the fact that the Wetterich equation (\ref{eq:twentyfour}) is an exact one-loop equation, for practical computations, one has to apply truncations in order to obtain
 a manageable theory space. The ground for the gravity-matter beta functions
 used in this section was  laid in \cite{Dona:2013qba,Dona:2014pla,Dona:2015tnf,Eichhorn:2017egq}.

\begin{figure}
    \centering
    \includegraphics[scale=0.7]{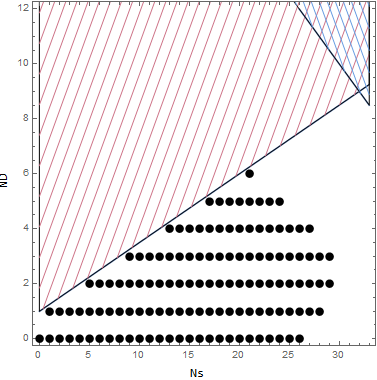}
    \caption{Dynamical matter degrees of freedom compatible with a gravitational fixed point with two relevant directions for $N_{V}=1$ (explicit values are listed in \autoref{AppendixA}), represented by black bullets. The shaded blue region represents a zone where a negative Newtons' fixed point takes place while the shaded red area contains points associated with $\xi_{\Tilde{\Lambda},2}<0$.}
    \label{fig:three}
\end{figure}

To perform the comparison between (\ref{eq:twelve}), (\ref{eq:sixteen}) and the functional renormalization group, equation (\ref{eq:twentyfour}), a matter sector is required in addition to the usual Einstein-Hilbert truncation. The matter content consists of $N_{s}$ scalar fields $\phi^{i}$, $N_{D}$ fermion fields $\psi^{i}$, and $N_{V}$ Abelian gauge fields $A^{i}_{\mu}$. In addition to this one has
the ghost and antighost fields $c$ and $\bar{c}$, all coupled to an external metric $g_{\mu\nu}$. The matter part of the action is given by,

\begin{eqnarray}
\Gamma_{matter}&=&S_{S}(\phi,g)+S_{D}(\psi,\bar{\psi},g)+S_{V}(A,c,\bar{c},g),\\
S_{S}(\phi,g)&=&\frac{1}{2}\int d^dx\sqrt{g}g^{\mu\nu}\sum^{N_{S}}_{i}\partial_{\mu}\phi^{i}\partial_{\nu}\phi^{i},\\
S_{D}(A,c,\bar{c},g)&=&i\int d^dx\sqrt{g}\sum^{N_{D}}_{i}\bar{\psi}_{i}{D\!\!\!\!/}\psi^{i},\\
S_{V}(A,c,\bar{c},g)&=&\frac{1}{4}\int d^dx \sqrt{g}\sum^{N_{V}}_{i=1}g^{\mu\nu}g^{\kappa\lambda}F^{i}_{\mu\kappa}F^{i}_{\nu\lambda}
+\frac{1}{2\xi}\int d^dx \sqrt{g}\sum^{N_{V}}_{i=1}\left(g^{\mu\nu}\nabla_{\mu}A^{i}_{\nu}\right)^2,\\
&&+\int d^dx \sqrt{g}\sum^{N_{V}}_{i=1}\bar{c}_{i}(-\nabla^2)c_{i},
\end{eqnarray}
where ${D\!\!\!\!/}=\gamma^{a}e^{\mu}_{a}\nabla_{\mu}$, with the orthonormal frame $e^{\mu}_{a}$ and where \textit{i} is a summation index over matter species. Adding to this action the contribution of the Einstein Hilbert truncation (using a type II cutoff), ghost and gauge fixing, the beta functions in four dimensions for the gravitational scale-dependent coupling are,

\begin{subequations}
\begin{align}
        \label{eq:twentyfiveone}
        \beta_{g}=&2g+\frac{g^2}{6\pi}\left(N_{S}+2N_{D}-4N_{V}-46\right),
        \\
        \label{eq:twentyfivetwo}
        \beta_{\lambda}=&-2\lambda+\frac{g}{4\pi}(N_{S}-4N_{D}+2N_{V}+2) \nonumber\\ 
        &+\frac{g\lambda}{6\pi}\left(N_{S}+2N_{D}-4N_{V}-16\right).
\end{align}
\end{subequations}
Here,~(\ref{eq:twentyfivetwo}) was expanded up to second order in $g$ and $\lambda$. The numbers $-46$, $+2$, and $-16$ describe the contribution of the ghost and metric sector, and they depend on the type of cut-off $\mathcal{R}_{k}$, which is used. \\
In the following analysis, the results from section \ref{sec:three} will
 be compared with the evolution of gravitational couplings using the FRGE. 
 To find suitable conditions for the relevant variables involved in the comparison, one shall vary the numbers of fields $N_{D}$, $N_{S}$, and $N_{V}$ such that the conditions encountered in the AS scenario are met.
 The flow equations~(\ref{eq:twentyfiveone}) and (\ref{eq:twentyfivetwo}) can be integrated analytically, and the dimensionful running version of  Newtons' coupling turns out to be
\begin{align}\label{eq:twentysix}
        G_{N}=\frac{G_{k}}{1+\frac{1}{2}\mathcal{C}_{1}k^2G_{k}},
\end{align}
where $G_{N}$ is the Newton constant measured in the deep infrared $k \rightarrow 0$. The dimensionful running version of the cosmological constant gives
\begin{align}\label{eq:twentyseven}
        \Lambda_{k}=&-\frac{\mathcal{C}_{2}}{\mathcal{C}_{3}\left(\mathcal{C}_{1}+\mathcal{C}_{3}\right)}\left(\frac{2+\left(\mathcal{C}_{1}+\mathcal{C}_{3}\right)k^2G_{k}}{G_{k}}\right)+\frac{\Lambda_{0}G_{N}\left(2+\mathcal{C}_{1}k^2G_{k}\right)^{1+\frac{\mathcal{C}_{1}}{C_{3}}}}{2^{1+\frac{\mathcal{C}_{3}}{\mathcal{C}_{1}}}G_{k}}
        \nonumber\\ 
        &+\frac{2^{-\frac{\mathcal{C}_{3}}{\mathcal{C}_{1}}}\mathcal{C}_{2}\left(2+\mathcal{C}_{1}k^2G_{k}\right)^{1+\frac{\mathcal{C}_{3}}{\mathcal{C}_{1}}}}{\mathcal{C}_{3}\left(\mathcal{C}_{1}+\mathcal{C}_{3}\right)G_{k}}.
\end{align}
In~(\ref{eq:twentysix}) and (\ref{eq:twentyseven}) we have defined 
\begin{subequations}\label{eq:twentyeight}
    \begin{align}
            \label{eq:twentyeight.one}
            \mathcal{C}_{1}=& \frac{1}{6\pi}\left(N_{S}+2N_{D}-4N_{V}-46\right),
            \\
            \label{eq:twentyeight.one}
            \mathcal{C}_{2}=& \frac{1}{4\pi}\left(N_{S}-4N_{D}+2N_{V}+2\right),
            \\
            \mathcal{C}_{3}=&\frac{1}{6\pi}\left(N_{S}+2N_{D}-4N_{V}-16\right).
    \end{align}
\end{subequations}
Consider the expansion around the quantity $k=(k^{'}+m_{0})/m_{0}$ (see~\autoref{fig:referencescale}). 
In the infrared the modified running $\Tilde{\Lambda}$ up to order ${k^{'}}^2$ reads, 
\begin{align}
    \label{eq:twentynine}
    \Tilde{\Lambda}_{k}=\Tilde{\Lambda}_{0}+m^4_{0}\mathcal{C}_{2}\cdot \left(\frac{k^{'}}{m_{0}}\right)+3\cdot2^{-\mathcal{C}_{1}}m^4_{0}\mathcal{C}_{2}\cdot \left(\frac{k^{'}}{m_{0}}\right)^2+\mathcal{O}\left(G_{0},\Lambda_{0}\right).
\end{align}
Since the crucial requirement of the method described in the two previous sections was $\xi_{\Tilde{\Lambda},2}>0$, 
the compatibility will be dictated by the sign of $\mathcal{C}_{2}$. 
To compare our result with those found in the AS program, the requirements that the gravitational parameters ($\xi_{\Tilde{\Lambda},1},\xi_{\Tilde{\Lambda},2}$) need to meet can be summarized as follows,
\begin{enumerate}
    \item \textit{Positive Newtons' fixed point $g^*>0$}: The low value ($k\lesssim
M_{pl}$,) of Newtons' gravitational coupling, is restricted by observations based on laboratory experiments at the scale $k_{lab}\backsimeq
 10^{-5} eV$.
    \item \textit{Relevant directions}: Insofar as the corresponding fixed points for the gravitational couplings of a pure gravity-theory have two relevant directions \cite{Dona:2013qba}, one expects that the addition of a small number of matter degrees of freedom does not change this behavior and the subsequent parametrization in theory space.
    \item \textit{Positive value of $\xi_{\Tilde{\Lambda},2}>0$}: As discussed before, the requirement to ensure that our model guarantees SSB at the level of the effective action needs $\lambda>0$ in (\ref{eq:sixteen:two}), or equivalently, $\mathcal{C}_{2}>0$ in (\ref{eq:twentynine}).
\end{enumerate}
The first two criteria were already pointed out in~\cite{Dona:2013qba}, and they are shown in \autoref{AppendixA} for different matter field configurations, while the third selection criterium is necessary for the validity in (\ref{eq:sixteen:two}). These conditions determine how many fields $N_S,\, N_D, \dots$ may be incorporated such that the proposed mechanism for SSB is in agreement with the requests of AS.  \autoref{fig:three} shows 
how the conditions above  allow to put restrictions 
on the field content 
 in the $N_{S}$-$N_{D}$-plane  for $N_{V}=1$. 
 Each black point indicates a viable model with SSB and an UV fixed point.

\newpage

\section{Summary and conclusion}
\label{sec:six}

In the present work, a novel mechanism for spontaneous symmetry breaking is 
suggested that circumvents the appearance 
of quadratic divergences by avoiding the breaking to take place
at the classical level. 
It is shown that SSB still can occur at the quantum level,
namely after setting the renormalization scale $k$ of the effective quantum action $\Gamma_k$.
As scale setting procedure, the VPS method is used.
This allows arriving at an optimized effective quantum action $\Gamma_{opt}$,
which, under certain conditions, produces SSB.

Despite the fact that the toy models of our study do not contain Yukawa, weak, or strong couplings,
the underlying mechanism can be expected to work also for models
containing these features.
It is shown that within this type of model, one can impose phenomenological conditions 
on $\Gamma_{opt}$. These conditions (in particular, the requirement $\xi_{\Tilde{\Lambda},2}>0$)
do then allow to put restrictions on the free parameters and the number of scalar, vector, and Dirac fields.
For the example of Asymptotically Safe quantum gravity coupled to matter,
it is shown that, for a given number of scalar fields, 
these conditions impose an upper and a lower bound on the number of Dirac fields, as shown in \autoref{fig:three}.

We further analyze to which extent the results depend
on the gauge choice, the truncation, and the shape of the cut-off function in \autoref{AppendixB}.
The inclusion of graviton and ghost anomalous dimension, 
as well as the anomalous dimension of the matter fields, derived in~\cite{Dona:2013qba},  
does not affect the findings discussed above. 

In a future study, we plan to perform an implementation  with all couplings necessary to arrive at
the Glashow-Weinberg-Salam model~\cite{Glashow:1961tr,Salam:1968rm,Weinberg:1967tq} coupled
to gravity.

\acknowledgments
The work of B.K. was supported by Fondecyt 1181694.

\newpage

\appendix
\section{Fixed points and relevant directions}\label{AppendixA}

\begin{table}[!hb]
\begin{footnotesize}
\begin{center}
\begin{tabular}{ |c|c|c|c|c|c|  }
 \hline
 \hline
 $N_{s}$ & $N_{D}$ & $g^*$ & $\lambda^*$ & $\theta_{1}$ & $\theta_{2}$\\
 \hline
 \hline
0 & 1 & 0.7891 & -0.0355 & 3.4679 & 1.8295\\ \hline
4 & 1 & 0.7874 & 0.0474 & 3.3244 & 1.9828\\ \hline
8 & 1 & 0.7712 & 0.1323 & 3.3392 & 2.1429\\ \hline
12 & 1 & 0.7985 & 0.2119 & 4.7225 & 1.6221\\ \hline
16 & 1 & 1.984 & 0.236 & 16.0381 & 4.2369\\ \hline
20 & 1 & 5.8534 & 0.1827 & 59.1008 & 15.8484\\ \hline
25 & 1 & 27.9783 & 0.0831 & 468.104 & 56.6011\\ \hline
30 & 1 & -7.4924 & -3.0458 & 4.3406 & 3.0051\\ \hline
35 & 1 & -4.3416 & -2.2382 & 4.2789 & 2.7502\\ \hline
0 & 2 & 0.9061 & -0.1388 & 3.6279 & 1.6733\\ \hline
4 & 2 & 0.9255 & -0.0513 & 3.5503 & 1.7274\\ \hline
8 & 2 & 0.9161 & 0.0457 & 3.5004 & 1.8615\\ \hline
12 & 2 & 0.8959 & 0.1407 & 3.8755 & 1.9107\\ \hline
16 & 2 & 1.0522 & 0.2183 & 7.004 & 1.1397\\ \hline
20 & 2 & 2.9317 & 0.2021 & 23.4523 & 5.4593\\ \hline
25 & 2 & 13.2672 & 0.0905 & 139.134 & 16.5161\\ \hline
30 & 2 & -6.5262 & -2.1898 & 4.6413 & 2.9661\\ \hline
0 & 4 & 1.2332 & -0.4412 & 3.8211 & 1.4862\\ \hline
4 & 4 & 1.3634 & -0.3722 & 3.8152 & 1.4125\\ \hline
8 & 4 & 1.4801 & -0.2736 & 3.8299 & 1.3405\\ \hline
12 & 4 & 1.5275 & -0.1386 & 3.9264 & 1.3165\\ \hline
16 & 4 & 1.4747 & 0.0126 & 4.1321 & 1.6424 \\ \hline
20 & 4 & 1.5462 & 0.1276 & 6.4885 & 1.4076\\ \hline
25 & 4 & 4.5202 & 0.0938 & 24.9725 & 2.0416\\ \hline
0 & 6 & 1.7796 & -0.9772 & 3.919 & 1.3963\\ \hline
4 & 6 & 2.2679 & -1.073 & 3.9343 & 1.2565\\ \hline
8 & 6 & 3.2689 & -1.3068 & 3.9642 & 1.0238\\ \hline
12 & 6 & 10.5941 & -3.5853 & 3.9829 & 0.3888\\ \hline
0 & 8 & 2.8677 & -2.096 & 3.9688 & 1.3603\\ \hline
4 & 8 & 5.0095 & -3.304 & 3.984 & 1.1922\\ \hline
8 & 8 & 63.0118 & -112.51 & 2.1179 & -1.0674\\ \hline
0 & 9 & 3.9567 & -3.2412 & 3.9837 & 1.3547\\ \hline
2 & 9 & 5.6446 & -4.423 & 3.99 & 1.2799\\ \hline
4 & 9 & 10.5569 & -7.9074 & 3.9964 & 1.1821\\ \hline
6 & 9 & 3.72\text{x}$10^7$ & -1087.061 & 1.25\text{x}$10^9$ & 6.22\text{x}$10^8$\\ \hline
0 & 10 & 6.1542 & -5.5762 & 3.9945 & 1.3547\\ \hline
2 & 10 & 11.888 & -10.3525 & 4 & 1.2801\\ \hline
4 & 10 & 712554.8395 & 490.2162 & 3.487\text{x}$10^6$ & 1.74\text{x}$10^6$\\ \hline
0 & 11 & 12.9664 & -12.8612 & 4.0024 & 1.3589 \\ 
\hline
\hline
\end{tabular}
\end{center}
\caption{
Selected gravitational fixed points and relevant directions for $N_{V}=1$ for type II cutoff, Feynman-de Donder gauge and one loop approximation. The first and second column indicate the matter content. The third and fourth column are the fixed points for the Newtons' and cosmological constant. The fifth and sixth column represents 
the negative value of the critical exponents.}\label{tabletwo}
\end{footnotesize}
\end{table}


\newpage

\section{Consistency of flow equations}\label{AppendixB}
Formulations of the FRGE require the inclusion of an IR regulator to ensure the integration of all degrees of freedom of fields possessing fluctuations of momenta higher than $k$. The choice of the arguments of the cut-off function $\mathcal{R}_{k}$ gives rise to diversity in the shape of the couplings' flow involved in the theory. However, physical results must remain independent of the selection of the shape and the corresponding endomorphism used in the cut-off function.
Thus,  one has to check whether the results obtained in this work are independent of this choice. 
Since the criteria for discriminating the result of the presented mechanism has to do with the sign of $\mathcal{C}2$, different truncations with various types of cut-off and expansion of the cosmological constant were investigating setting $N_{S}=2$, $N_{V}=1$ and the number of Dirac fields being a number between 2 and 9, where we know the model works while fulfilling the conditions required by the AS. The results  presented in~\autoref{tableone} confirm that this analysis is robust under changes in the truncation procedure.

\begin{table}[hb!]
\begin{center}
\begin{tabular}{ |c|c|c|c|c|  }
 \hline
 \hline
 Ref & Truncation & Gauge & Specifics & sgn($\xi_{\Tilde{\Lambda},2}$)\\
 \hline
 \hline
~\cite{Dona:2013qba}& \small{EH with SM matter} & \small{$\alpha=\beta=1$} & \small{type Ia cutoff lowest order in $\Lambda$} & \small{Positive}\\ \hline
~\cite{Dona:2013qba} & \small{EH with SM matter} & \small{$\alpha=\beta=1$} & \small{type Ia cutoff first order in $\Lambda$} & \small{Positive}\\ \hline
~\cite{Dona:2013qba} & \small{EH with SM matter} & \small{$\alpha=\beta=1$} & \small{type Ib cutoff lowest order in $\Lambda$} & \small{Positive} \\ \hline
~\cite{Dona:2013qba} & \small{EH with SM matter} & \small{$\alpha=\beta=1$} & \small{type Ib cutoff first order in $\Lambda$} & \small{Positive} \\ \hline
~\cite{Dona:2013qba} & \small{EH with SM matter} & \small{$\alpha=\beta=1$} & \small{type II cutoff lowest order in $\Lambda$} & \small{Positive} \\ \hline
~\cite{Dona:2013qba} & \small{EH with SM matter} & \small{$\alpha=\beta=1$} & \small{type II cutoff first order in $\Lambda$} & \small{Positive} \\ \hline
~\cite{Eichhorn:2017lry}& \small{EH with SM matter} & \small{$\alpha=0,\beta=1$} & \small{type II cutoff lowest order in $\Lambda$} & \small{Positive} \\ \hline
~\cite{Eichhorn:2017lry} & \small{EH with SM matter} & \small{$\alpha=0,\beta=1$} & \small{type II cutoff first order in $\Lambda$} & \small{Positive} \\ \hline
~\cite{Alkofer:2018fxj}& \small{$f(R)$ to $R^9$ with SM matter} & \small{$\alpha=0,\beta=-\infty$} & \small{type I cutoff lower order in $\Lambda$} & \small{Positive} \\ \hline
~\cite{Alkofer:2018fxj}& \small{$f(R)$ to $R^9$ with SM matter} &\small{$\alpha=0,\beta=-\infty$} & \small{type I cutoff first order in $\Lambda$} & \small{Positive} \\ \hline
~\cite{Alkofer:2018fxj}& \small{$f(R)$ to $R^9$ with SM matter} &\small{$\alpha=0,\beta=-\infty$} & \small{type II cutoff lower order in $\Lambda$} & \small{Positive} \\ \hline
~\cite{Alkofer:2018fxj}& \small{$f(R)$ to $R^9$ with SM matter} &\small{$\alpha=0,\beta=-\infty$} & \small{type II cutoff first order in $\Lambda$} & \small{Positive}\\
\hline
\hline
\end{tabular}
\end{center}
\caption{ 
The compatibility of the result obtained in (\ref{eq:fifteen}) is investigated for various studies 
of the gravitational RG flow in the presence of SM massless matter fields minimally coupled to an external metric.
In the third column, different choices of the gauge parameters  $\alpha$ and $\beta$ are explored. 
 The information exhibited in the column "specifies" contains the choice of the covariant differential operator used as the argument of the cutoff function and the expansion of the cosmological constant, explained in the references of the first column. The fifth column gives the sign of the relevant parameter involved in the process of SSB in (\ref{eq:sixteen:two})}\label{tableone}
\end{table}

Note that the characterization of $\mathcal{C}_{2}$ is robust
under changes of the gauge choice~\cite{Gies:2015tca}.
Due to the structure of the $\mathcal{C}_{2}$ term in (\ref{eq:twentyeight.one}), the sign of $\xi_{\Tilde{\Lambda},2}$ is also independent of how many scalar fields are incorporated while keeping $N_{D}$ at some fixed number.

Most of the findings presented in this appendix might seem evident, but some issues appear when one gets solutions of the FRGE with one or the other of the kinetic operator. In particular, \cite{Codello:2008vh,Dona:2012am,Alkofer:2018fxj} shows that the spectrum of $\nabla^2$ (type I cut-off) and $\nabla^2-\frac{R}{4}$ (type II cut-off) may turn out in ambiguities in the sign of the fermionic contribution to the running of Newtons' constant. In other words, the background-field dependence of $\mathcal{R}_{k}$ can alter results in the background approximation of physical observables. 
Since only type II cut-off gives the sign according to the infrared observation of $G$. 
This result has been corroborated by employing a completely independent method to evaluate the r.h.s. of (\ref{eq:twentyfour})~\cite{Dona:2012am}. However, the SSB discussed in this paper takes place 
independent of this sign and independent of the gauge choices.



\end{document}